\documentclass[%
reprint,
superscriptaddress,
showpacs,preprintnumbers,
 amsmath,amssymb,
 aps,
]{revtex4-1}

\usepackage{graphicx}
\usepackage{dcolumn}
\usepackage{bm}

\begin{document}

\preprint{}

\date{}
 
\title{Subnanotesla quantum-interference magnetometry with a single spin in diamond}

\author{A. Angerer}
\author{T. N{\"o}bauer}
 \altaffiliation[Current address: ]{Research Institute of Molecular Pathology and Department of Structural and Computational Biology, University of Vienna, Dr. Bohr-Gasse 7, 1030 Vienna, Austria}
\author{G. Wachter}
\affiliation{Vienna Center for Quantum Science and Technology, Atominstitut, TU Wien, Stadionallee 2, 1020 Vienna}
\author{M. Markham}
\author{A. Stacey}
 \altaffiliation[Current address: ]{Centre for Quantum Computation and Communciation Technology, School of Physics, University of Melbourne, Parkville, Melbourne, VIC, Australia.}
\affiliation{Element Six Ltd., Global Innovation Centre, Fermi Avenue, Harwell Oxford, Didcot, OX11 0QR, UK}%
\author{J. Majer}
\author{J. Schmiedmayer}
\author{M. Trupke}
\email{All correspondence should be addressed to  michael.trupke@tuwien.ac.at}
\affiliation{Vienna Center for Quantum Science and Technology, Atominstitut, TU Wien, Stadionallee 2, 1020 Vienna}

\date{\today}

\begin{abstract}
We demonstrate a magnetometry technique using nitrogen-vacancy centres in diamond which makes use of coherent two-photon transitions. We find that the sensitivity to magnetic fields can be significantly improved in isotopically purified diamond. Furthermore, the long-term stability of magnetic field measurements is significantly enhanced, thereby reducing the minimum detectable field variations for both quasi-static and periodic fields. The method is useful both for sensing applications and as a spin qubit manipulation technique.
\end{abstract}

\maketitle 
Quantum interference in spin-1 systems has played a central role for precision measurements  and tests of quantum mechanics using ensembles of atoms and molecules \cite{Hellmuth1987,Hudson2012}. Coherent quantum control of three-level systems has furthermore been central to schemes for the coherent generation of single photons and the storage of light \cite{Kuhn2002, Fleischhauer2005}. Solid-state defects have recently emerged as powerful systems for applications in sensing, quantum information and quantum communication due to the discovery of emitters with long spin-coherence times in a multitude of crystalline hosts \cite{Doherty2013, Weber2010}. Single defects are particularly promising for nanoscale sensing applications as they promise unrivalled sensitivity-to-size ratios. The nitrogen vacancy (NV) defect in diamond, as well as several defects in silicon carbide, have shown particular promise as carriers of quantum information or as a sensors in bulk and nanocrystalline systems \cite{Dolde2011,Neumann2013,Kucsko2013, Rondin2014, Christle2015, Widmann2015, Fuchs2015, Knowles2014, Zu2014}.

The electronic spin-1 ground state of the NV centre can be subdivided into the qubit subspaces $\lbrace|0\rangle,|\pm 1\rangle\rbrace$ and $\lbrace|-1\rangle,|+1\rangle\rbrace$. We demonstrate the characterisation and manipulation of a qubit in the $\lbrace|+1\rangle ,\, |-1\rangle\rbrace$-subspace and make use of coherent interference effects to reduce the spin-1 system to an effective two-level system \cite{Shore1998,Fang2013}. We show that, under application of a modest magnetic bias field, this qubit displays significantly improved coherence times compared to the traditionally used $\lbrace|0\rangle ,\, |\pm 1\rangle\rbrace$-systems. Our measurements indicate that the reduction in dephasing rates is brought about by a reduced sensitivity to strain and transverse magnetic field fluctuations. These improvements in turn lead to significantly higher sensitivity for measurements of both arbitrary and periodic magnetic fields. The manipulation methods used herein furthermore enable efficient control over the NV qutrit manifold, which have the potential to significantly reduce the physical requirements for large-scale quantum information systems \cite{Lanyon2009}.
We anticipate that our findings will be beneficial for the application of NV centres as sensors and in quantum technology.

\begin{figure}[t!]
\centering
\includegraphics[width=\columnwidth]{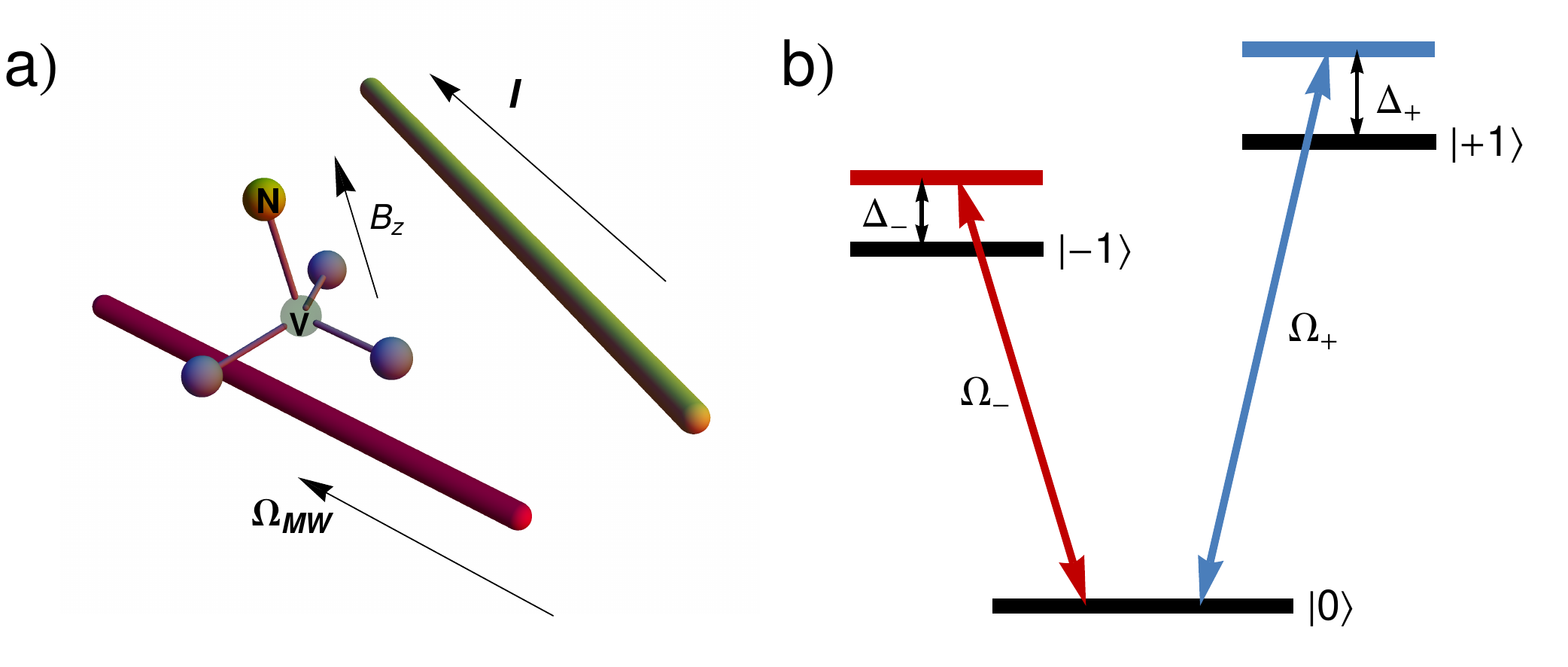}
\caption{a) The NV centre is driven with microwave fields $\Omega_{MW}$ brought to the sample using an external microwire. A wire perpendicular to the axis of the NV centre carries currents $I$ used to create the arbitrary and periodic magnetic fields. b) Level scheme of the spin-1 ground state, shown with the main detuning frequencies $\Delta_{+,-}$ and driving fields $\Omega_{+,-}$ used in the measurements.}\label{fig:overFig}
\end{figure}
For our measurements we use a single-crystal artificial diamond created by chemical vapour deposition. An epitaxially grown layer of $50\,\mu$m thickness, isotopically purified to 99.999\% $^{12}$C, hosts the single NV centres \cite{Ishikawa2012}. The experimental apparatus used for the measurements is a standard home-built confocal microscope setup. Optical excitation is provided by an intensity-stabilised diode-pumped solid-state laser at $532\,$nm. We monitor the fluorescence of the NV centres using an oil-immersion microscope objective with a numerical aperture of 1.42, and photons are collected through a dichroic mirror using two single-photon counting modules. Microwave fields were applied through a $100\,\mu$m wire gold spanned over the diamond surface [Fig. \ref{fig:overFig} a)].

We manipulate the NV centre spin state using four distinct microwave fields, two resonant with the $|0\rangle \leftrightarrow |\pm1\rangle$ transitions, and two fields which are detuned relative to the transition resonances by $\Delta_{+,-}$ [Fig. \ref{fig:overFig} b)]. The amplitudes of these four fields give full control over the state of the spin-1 system. The Hamiltonian of the system is given by
\begin{equation}
\hat{H}=\left[\begin{array}{ccc}
0			& \Omega_{+}						& 0				\\ 
\Omega_{+}	& 2\Delta_{+}	& \Omega_{-}	\\ 
0			& \Omega_{-}	& 2(\Delta_{+}-\Delta_{-})
\end{array}\right],
\end{equation}\label{Hamiltonian}
in a frame rotating at the frequency $\omega_{rot}=\Delta_+-\Delta_-$ \cite{Shore1998}. $\Omega_{+,-}$ are the Rabi frequencies of the driving fields detuned by $\Delta_{+,-}$. These fields can create an effective coupling between the states $|+1\rangle \leftrightarrow |-1\rangle$, which we can use to characterise the dephasing properties of this system. We apply a bias magnetic field of $30\,$mT and can therefore neglect small transverse magnetic fields and strain components \cite{Doherty2013}.

We first measured the spin coherence lifetime in the $\lbrace|+1\rangle ,\, |-1\rangle\rbrace$ qubit system. For this purpose, the NV centre is initialized in the state $|0\rangle$, after which a resonant $\pi$-pulse transfers the population into $|-1\rangle$. From here, a $\pi/2$-pulse performed with both $\Omega_{+}$ and $\Omega_{-}$ brings the system into the superposition $(|+1\rangle+|-1\rangle)/\sqrt{2}$. The driving amplitudes are both controlled with Gaussian envelopes, and the $\Omega_+$ pulse is delayed with respect to the $\Omega_-$ pulse. This sequence is a stimulated Raman adiabatic passage (STIRAP), as used for atomic samples in the optical domain \cite{Shore1998}. We interrupt the sequence halfway through, at the point where the desired superposition is achieved.  After a time delay the electron is transferred further into $|+1\rangle$ using the second half of the STIRAP sequence.  Subsequent measurements of the population in the three eigenstates of the undriven system yield a spin lifetime $T_2^*=102\,\mu$s, more than twice the value measured for the $\lbrace|0\rangle,|-1\rangle\rbrace$-qubit (40$\,\mu$s).

 To examine the matter further, we inspect the ground state Hamiltonian of the NV centre,\\
$\hat{H}_{GS}=$
\begin{equation}
{\fontsize{10}{12} 
\left( \begin{array}{ccc}
0 & g \mu_B \beta_{\perp} & g \mu_B\beta^*_{\perp} \\ 
g \mu_B\beta^*_{\perp} & D_g+d_{\parallel} E_z+g \mu_B B_z & d_{\perp} (E_x+i E_y) \\ 
g \mu_B\beta_{\perp}  & d_{\perp} (E_x-i E_y) & D_g+d_{\parallel} E_z-g \mu_B B_z
\end{array} \right).
}
\end{equation}\label{eq:gsHam}

Here $D_g\sim 2.87\,$GHz is the ground-state splitting, while $d_{\perp}=0.17\,$Hz m/V and $d_{\parallel}=35 \times 10^{-3}\,$Hz m/V are the perpendicular and parallel electric dipole components, respectively. $\beta_{\perp}=(i B_x-B_y)/\sqrt{2}$ collects the transverse components of the magnetic field vector, with $\beta_{\perp}^*$ its complex conjugate. $B_j$ and $E_j$, with $j=x,y,z$, are magnetic fields or strain applied in the three spatial directions. We now examine the effect of these components individually. Transverse magnetic fields lead to an approximately quadratic dependence of the transition frequencies for $B_{x,y}\ll B_z$. However the dependence is $\sim 5$ times smaller in the $|\pm 1\rangle$ system compared to the $|0\rangle \leftrightarrow |\pm 1\rangle$ transitions at the bias field of $30\,$mT, since the shift caused by transverse magnetic fields is positive for both upper eigenstates, but negative for the lowest state.

The effect of strain (or electric field fluctuations) $E_z$ along the axis of the defect can be seen by inspection of the Hamiltonian to be a linear frequency shift for $|0\rangle \leftrightarrow |\pm 1\rangle$ transitions, while causing no shift for the $|-1\rangle \leftrightarrow |+1\rangle$ transition. The energies of the $|\pm 1\rangle$ eigenstates at zero strain, $D_g\pm g\mu_B B_z$ are shifted to $D_g\pm \sqrt{(g\mu_B B_z)^2 + d_{\perp}^2 (E_x^2+E_y^2)}$ under transverse strain. Such perturbations will therefore cause a frequency shift with twice the magnitude for the $|\pm 1\rangle$ system. However, due to the addition in quadrature, their effect is strongly suppressed in both manifolds by application of a bias field $B_z$.

The improvement was observed for all NV centres examined in the isotope-purified portion of the sample, but no improvement was found for NV centres in the portion with natural isotopic abundance. This distinction leads us to conclude that the that the longer coherence times measured for the $|\pm 1\rangle$ may be brought about not by a better isolation from bath spins, but by a weaker dependence on environmental perturbations other than the measured magnetic field along the NV axis.

\begin{figure*}[t!]
\centering
\includegraphics[width=2\columnwidth]{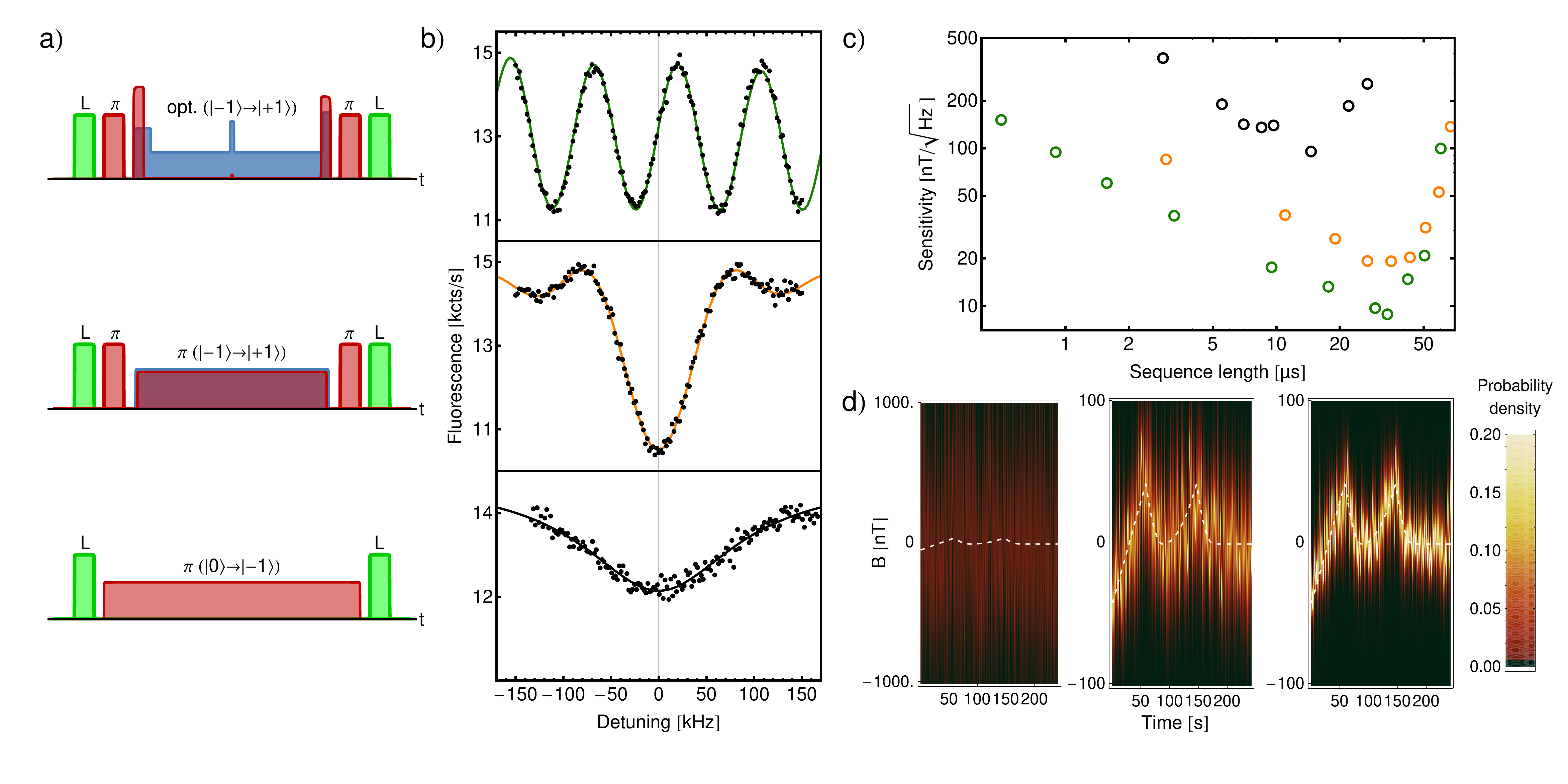} 
\caption{Sensing of arbitrary magnetic fields. a) Pulse sequence for pulsed ODMR in the $\lbrace|0\rangle,|-1\rangle\rbrace$-manifold (bottom),  a continuous two-tone $|-1\rangle \rightarrow |+1\rangle $ transfer pulse (middle) and a numerically optimised time-dependent pulse (top). The red, blue and green areas represent the $\Omega_+$, $\Omega_-$ and laser pulses applied. The $\pi$-pulses are performed on resonance. b) Detuning-dependent population for the sequences shown in a). c) Sensitivity as a function of sequence length for pulsed optically detected magnetic resonance (ODMR) (black), continuous transfer (orange) and optimised pulses (green). d) Measurement of the same arbitrary sub-$\mu$T time-dependent magnetic field (dashed white line), compared for pulsed ODMR on the $|0\rangle \rightarrow |-1\rangle $ transition (left), continuous transfer on the $|-1\rangle \rightarrow |+1\rangle $ transition (middle), and numerically optimised transfer on the $|-1\rangle \rightarrow |+1\rangle $ transition (right). The colour scale indicates the probability density for the magnetic field value, extracted from the NV fluorescence for one second of integration time.}\label{dcFig}
\end{figure*}
\begin{figure*}[t!]
\centering
\includegraphics[width=2\columnwidth]{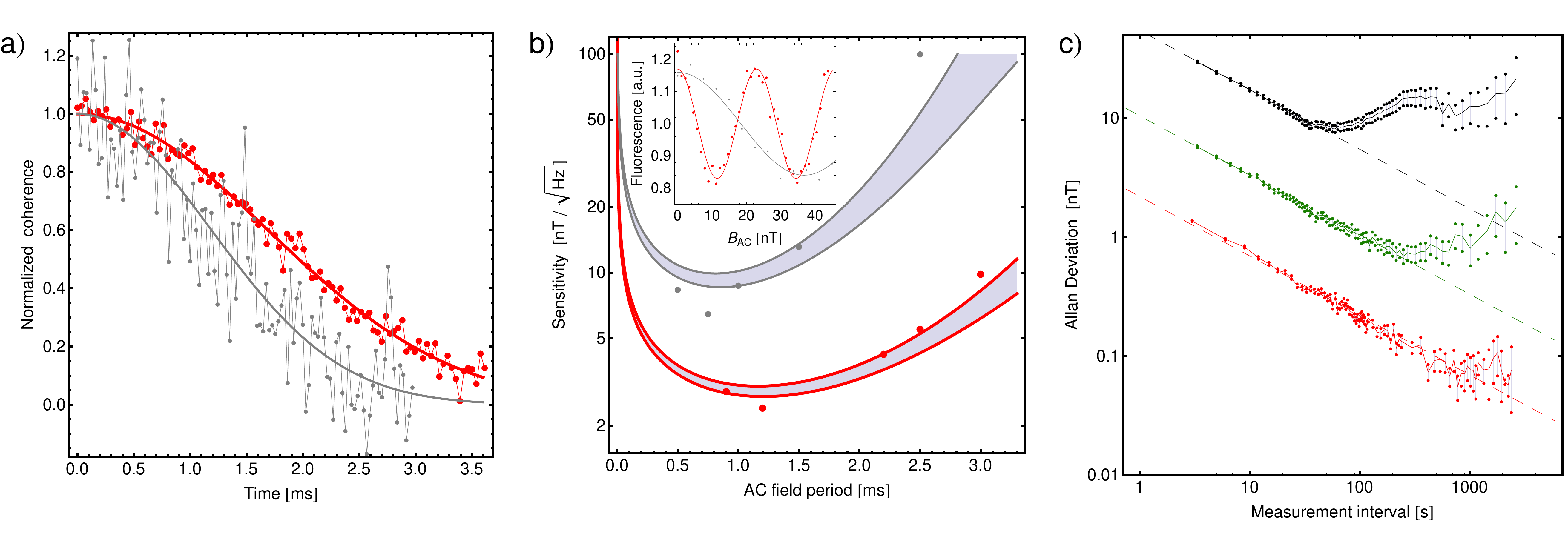} 
\caption{a) Comparison of $T_2$ coherence time measurements for the $|\pm 1\rangle$-qubit (red, $T_2=(2.36 \pm 0.09)\,$ms) and the $(|0\rangle,|-1\rangle)$ qubit (grey, $T_2(0,-1)=(1.66\pm 0.16)\,$ms). b) Sensitivity to oscillating magnetic fields for both systems versus field oscillation period. The shaded areas indicate the sensitivity expected from the $T_2$ coherence time measurements, broadened by their uncertainty. Inset: Sinusoidal dependence of the fluorescence on the magnitude of an applied AC magnetic field for the two systems, recorded for phase accumulation times $T_2 /2$. c) Allan deviation for three measurement types: DC magnetometry with pulsed ODMR in the $(|0\rangle,|-1\rangle)$ basis (black); DC magnetometry with optimized pulses in the $(|-1\rangle,|+1\rangle)$ basis (green); and AC magnetometry in the $(|0\rangle,|-1\rangle)$ basis (red). The dashed lines show the $1/\sqrt{t}$ dependence expected for purely shot-noise limited measurements.}\label{acFig}
\end{figure*}

 
We now describe how to make use of the improved coherence lifetime in the $|\pm 1\rangle$ system to measure arbitrary magnetic fields. The measurement principle is related to that used in the determination of the dephasing rate, as well as in previous magnetometry measurements using the $\lbrace|0\rangle,|1\rangle\rbrace$ and $|\pm1\rangle$ systems \cite{Fang2013}. 
As a basis for comparison, we measure the magnetic field sensitivity for the $|0\rangle \rightarrow |-1\rangle$ transition using the pulsed electron spin resonance (ESR) method [Fig. \ref{acFig} a)] \cite{Rondin2014}. The sensitivity can be found by recording the ESR spectrum of the transition [Fig. \ref{acFig} b)] and determining the largest gradient $G_{max}$, so that 
\begin{equation}
\Delta B_{DC}^{min}=\sqrt{\frac{\tau_{DC}}{N}}\times \frac{1}{G_{max}}\, ,
\end{equation}\label{dcSens}
where $\tau_{DC}$ is the duration of the measurement sequence and $N$ is the number of photons collected on average per sequence at the microwave detuning corresponding to the maximum gradient.

We compare the sensitivity of the pulsed ESR measurement to a similar measurement in which the spin is coherently transferred along $|-1\rangle \rightarrow |+1\rangle$. In order to make optimal uses of the extended coherence time of the $|\pm 1\rangle$ qubit, we transfer the system to $|-1\rangle $ immediately after initialisation. The transition amplitudes for the effective two-level system are sensitive to the relative detuning $\Delta_{tot}=\Delta_{+}-\Delta_{-}-2g\mu_B\delta B_z$, with $g\mu_B=28\,$Hz/nT, and are therefore twice as sensitive as the simple qubit transition. The longer coherence time of the $\lbrace|+1\rangle ,\, |-1\rangle\rbrace$-qubit allows for longer phase accumulation times, further increasing the sensitivity.

We then measured the sensitivity for a more sophisticated sequence. For this measurement the spin is prepared in $|-1\rangle$, then transferred into a superposition $(\alpha|+1\rangle+e^{i\phi}\beta|-1\rangle)/\sqrt{\alpha^2+\beta^2}$, before being transferred further to the state $|+1\rangle$. In the experiment we use a numerically optimised pulse sequence provided by a genetic algorithm \cite{Rohringer2013}. The simulation takes into account the rotation of the state throughout the sequence to maximise the dependence on the magnetic field of the population in $|-1\rangle$ after the sequence. The results show a reduction in the minimum detectable field, as shown in Fig. \ref{dcFig} c).  Finally, we show the effect of this improvement on the measurement of an actual varying magnetic field. The field is applied using a wire carrying a time-varying current $I$, which is spanned perpendicularly to the NV centre axis, thereby creating a magnetic field along the NV axis as depicted in Fig. \ref{fig:overFig} a). The field at each current value is measured for one second, allowing us to extract a statistical uncertainty from the distribution of the recorded fluorescence counts.  The greater sensitivity markedly improves this practical measurement, as shown in Fig. \ref{dcFig} d).
The techniques described above make possible the application of rephasing pulse sequences known from nuclear magnetic resonance experiments and previous experiments with solid-state defects \cite{Balasubramanian2009}. To measure the rephased coherence time $T_2$, we perform a similar sequence as for the $T_2^*$ measurement described above. This time however, after half the time delay, we rotate the state about the mean Bloch vector to perform a Carr-Purcell-Meiboom-Gill (CPMG) sequence. This rotation is achieved using synchronous $\Omega_{+,-}$ pulses with Gaussian envelopes.  Once again, using the $|\pm 1\rangle$ states, the coherence time of the NV centre is increased to $T_2(\pm 1)=(2.36\pm 0.09)\,$ms , compared to $T_2(0,-1)=(1.66\pm 0.16)\,$ms [Fig. \ref{acFig} a)]. While this increase is not as dramatic as for the $T_2^*$ values, it nonetheless represents a significant improvement, enabling improved measurements of oscillating fields: The long phase accumulation time warranted by the rephasing sequence can be used to perform highly sensitive magnetic field measurements, albeit only for known phases and frequencies of the magnetic field amplitude \cite{Balasubramanian2009}. The smallest oscillating field that can be measured using this method is given by 
\begin{equation}
\Delta B_{AC}^{min}=\sqrt{\frac{1}{N \tau_{AC}}}\times \frac{1}{2 C_0 \exp[-(\tau_{AC}/T_2)^p]\delta_S g\mu_B}\,.
\end{equation}\label{acSens}
Here $N$ is the number of photons collected in a single acquisition and $\tau_{AC}$ is the oscillation period of the field. $C_0$ is the normalized fluorescence contrast for the two states and $p$ describes the shape of the coherence decay. Our measurements yield $p=(2.03\pm0.38)$ for the $\lbrace|0\rangle,|-1\rangle\rbrace$ system and $p=(2.02\pm0.1)$ for the $|\pm 1\rangle$ system. The difference in magnetic quantum number $\delta_S=1$ for the $\lbrace|0\rangle ,\, |-1\rangle\rbrace$ and $\delta_S=2$ for the $|\pm 1\rangle$ system.

Once again, the factor-of-two increase in magnetic field dependence and the improved coherence time markedly reduce the minimum detectable field (see Fig. \ref{acFig} b)).  Measurements in the $\lbrace|0\rangle,|\pm 1\rangle\rbrace$-system give a sensitivity of $6\,$nT/Hz$^{1/2}$, while significantly smaller changes in the field modulation amplitude of $2.2\,$nT/Hz$^{1/2}$ were detected using the $|\pm 1\rangle$ system. The improvement is even more significant for longer field oscillation periods, as shown in Fig. \ref{acFig} b).

 A further important feature of the measurement techniques demonstrated here is the long-term stability given by the common-mode response of the $|\pm 1\rangle$ levels to extraneous noise sources. Fig. \ref{acFig} c) shows the Allan deviation for DC and AC measurements. The measurement of static fields in the $\lbrace|0\rangle,|-1\rangle\rbrace$-basis reaches its floor of $10\,$nT for time intervals of approximately one minute. This timescale is consistent with temperature fluctuations observed in the laboratory, which affect both the NV centre directly, as well as the permanent magnet used to apply the bias field. The same measurement performed using the optimised pulses for the $|\pm 1\rangle$ qubit remains stable for intervals which are an order of magnitude longer, and reaches a minimum deviation of $\sim 800\,$pT. We presume this sensitivity to be limited by actual fluctuations of the axial magnetic field around the sample. Neither of these mechanisms have a discernible effect on AC magnetometry using the $|\pm 1\rangle$ qubit, which displays shot-noise limited behaviour over several hours. The Allan deviation of this measurement reaches $\sim 50\,$pT for intervals of approximately half an hour, limited only by the statistics and duration of our measurement.
 
To conclude, the use of the full spin-1 manifold of the NV ground state gives access to numerous possibilities for storing and manipulating quantum states. The increase in coherence time given by encoding qubit states into the states $|\pm 1\rangle$ when using isotopically purified diamond will be beneficial for quantum information applications, and leads to significant improvements in the sensitivity to static and modulated magnetic fields. The sensitivity is limited by the shot noise on the detected photons, and can be improved further by implementing existing micro-optical elements \cite{Li2014}, or by using sparse NV ensembles. Potential improvements also include spin to charge conversion or photoelectron detection \cite{Shields2015,Bourgeois2015}.

 We furthermore note that the use of two-photon manipulation will enable the use of crossed-wire grids for the manipulation of arrays of NV centres, leading to a quadratic reduction in the number of connections required for full control of qubit grids in large-scale quantum information systems \cite{Zu2014,Derntl2014,Arroyo2014,Nemoto2014}. Given these advantages, we expect the techniques demonstrated in this work to be of broad use for applications of solid-state defects as sensors and carriers of quantum information.
 
\textbf{Acknowledgements} We thank M. Bichler and A. Pelczar for assistance with custom electronics, and M. Zens for assistance in the AC measurements. Financial support was provided by the FFG program PLATON-NAP III, The TU Wien program Innovative Projekte, the EU FP7 program DIAMANT and the FWF graduate school programs CoQuS and Solids4Fun.
\\

\end{document}